\documentclass[%
 reprint,
superscriptaddress,
 amsmath,amssymb,
 aps,
prb,
]{revtex4-2}

\usepackage{color}
\usepackage{graphicx}
\usepackage{dcolumn}
\usepackage{bm}
\usepackage{hyperref}
\hypersetup{colorlinks=true,linkcolor=blue,citecolor=blue,urlcolor=blue}

\usepackage[T1]{fontenc} 
\usepackage{newtxtext}   
\usepackage{newtxmath}   
\usepackage{upgreek}

\begin{document}

\preprint{APS/123-QED}

\title{Pressure tunable magnetic skyrmion phase in Co$_8$Zn$_8$Mn$_4$ single crystals}  

\author{Zhun Li}
\thanks{These authors contributed equally to this work. }
\affiliation{College of Physics \& Center of Quantum Materials and Devices, Chongqing University, Chongqing 401331, China}
\affiliation{Institute of High Energy Physics, Chinese Academy of Sciences (CAS), Beijing 100049, China}
\affiliation{Spallation Neutron Source Science Center, Dongguan 523803, China}   
\affiliation{Guangdong Provincial Key Laboratory of Extreme Conditions, Dongguan, 523803, China}

\author{Xinrun Mi}
\thanks{These authors contributed equally to this work. }
\affiliation{College of Physics \& Center of Quantum Materials and Devices, Chongqing University, Chongqing 401331, China}

\author{Xinming Wang}
\affiliation{Institute of High Energy Physics, Chinese Academy of Sciences (CAS), Beijing 100049, China}
\affiliation{Spallation Neutron Source Science Center, Dongguan 523803, China}   
\affiliation{Guangdong Provincial Key Laboratory of Extreme Conditions, Dongguan, 523803, China}

\author{Jian Lyu}
\affiliation{Institute of High Energy Physics, Chinese Academy of Sciences (CAS), Beijing 100049, China}
\affiliation{Spallation Neutron Source Science Center, Dongguan 523803, China}   
\affiliation{Guangdong Provincial Key Laboratory of Extreme Conditions, Dongguan, 523803, China}

\author{Na Su}
\affiliation{College of Physics \& Center of Quantum Materials and Devices, Chongqing University, Chongqing 401331, China}

\author{Aifeng Wang}
\affiliation{College of Physics \& Center of Quantum Materials and Devices, Chongqing University, Chongqing 401331, China}
 
\author{Yisheng Chai}
\affiliation{College of Physics \& Center of Quantum Materials and Devices, Chongqing University, Chongqing 401331, China}

\author{Bao Yuan}
\affiliation{Institute of High Energy Physics, Chinese Academy of Sciences (CAS), Beijing 100049, China}
\affiliation{Spallation Neutron Source Science Center, Dongguan 523803, China}   
\affiliation{Guangdong Provincial Key Laboratory of Extreme Conditions, Dongguan, 523803, China}

\author{Wanju Luo}
\affiliation{Institute of High Energy Physics, Chinese Academy of Sciences (CAS), Beijing 100049, China}
\affiliation{Spallation Neutron Source Science Center, Dongguan 523803, China}   
\affiliation{Guangdong Provincial Key Laboratory of Extreme Conditions, Dongguan, 523803, China}

\author{Hui Cheng}
\affiliation{Institute of High Energy Physics, Chinese Academy of Sciences (CAS), Beijing 100049, China}
\affiliation{Spallation Neutron Source Science Center, Dongguan 523803, China}   
\affiliation{Guangdong Provincial Key Laboratory of Extreme Conditions, Dongguan, 523803, China}

\author{Jianxiang Gao}
\affiliation{Neutron Scattering Laboratory, Department of Nuclear Physics, China Institute of Atomic Energy, Beijing, 102413, Peoples R China}

\author{Hongliang Wang}
\affiliation{Neutron Scattering Laboratory, Department of Nuclear Physics, China Institute of Atomic Energy, Beijing, 102413, Peoples R China}

\author{Lijie Hao}
\affiliation{Neutron Scattering Laboratory, Department of Nuclear Physics, China Institute of Atomic Energy, Beijing, 102413, Peoples R China}

 \author{Mingquan He}
\email{mingquan.he@cqu.edu.cn}
\affiliation{College of Physics \& Center of Quantum Materials and Devices, Chongqing University, Chongqing 401331, China}

\author{Junying Shen}
\email{shenjy@ihep.ac.cn}
\affiliation{Institute of High Energy Physics, Chinese Academy of Sciences (CAS), Beijing 100049, China}
\affiliation{Spallation Neutron Source Science Center, Dongguan 523803, China} 
\affiliation{Guangdong Provincial Key Laboratory of Extreme Conditions, Dongguan, 523803, China}

\author{Young Sun}
\email{youngsun@cqu.edu.cn}
\affiliation{College of Physics \& Center of Quantum Materials and Devices, Chongqing University, Chongqing 401331, China}

\author{Xin Tong}
\email{tongx@ihep.ac.cn}
\affiliation{Institute of High Energy Physics, Chinese Academy of Sciences (CAS), Beijing 100049, China}
\affiliation{Spallation Neutron Source Science Center, Dongguan 523803, China}
\affiliation{Guangdong Provincial Key Laboratory of Extreme Conditions, Dongguan, 523803, China}

\date{\today}

\begin{abstract}
In a magnetic skyrmion phase, magnetic moments form vortex-like topological textures which are of both fundamental and industrial interests. In $\beta$-Mn-type Co-Zn-Mn alloys, chrial magnetic skyrmions emerge above room temperature, providing a unique system for studying the skrymion physics and exploring spintronics applications. However, the magnetic skyrmion phase is typically confined in a narrow and limited temperature ($T$) and magnetic field ($H$) range. Here, we demonstrate that hydrostatic pressure can expand the skyrmion phase in the $T-H$ phase diagram of single-crystalline Co$_8$Zn$_8$Mn$_4$. At ambient pressure, signatures of skyrmions are seen within $T\sim302-308$  K and $H\sim50-100$ Oe. Applying a moderate pressure of 6 kbar extends this range to $T\sim300-310$  K and $H\sim50-150$ Oe. However, further escalation of pressure to 10 kbar results in a slight contraction of the skyrmion phase. These findings underscore the sensitivity of the skyrmion phase in Co$_8$Zn$_8$Mn$_4$ to external pressures, and hint at the potential of strain engineering, particularly in $\beta$-Mn-type Co-Zn-Mn thin films, as a promising avenue to customize the skyrmion phase.  



\end{abstract}

\maketitle

\section{Introduction}

Magnetic skyrmions, distinguished by their unique vortex-like spin configurations carrying integer topological charges, have become a forefront in the study of condensed matter physics and material science  \cite{Bogdanov1989,muhlbauer2009skyrmion,yu2010real}. The nontrivial spin textures of skyrmions are fascinating both from a fundamental physics perspective and for their potential technological applications. Fundamentally, magnetic skyrmions can arise through diverse mechanisms, such as the interplay between ferromagnetic exchange, Dzyaloshinskii-Moriya (DM), Ruderman-Kittel-Kasuya-Yosida (RKKY), and magnetic dipolar interactions \cite{muhlbauer2009skyrmion,yu2010real,yu2014biskyrmion,wang2016centrosymmetric,kurumaji2019skyrmion,hirschberger2019skyrmion}. From a technological standpoint, skyrmions stand out for their emergent electromagnetic properties and the ability to be manipulated with low current densities, making them highly efficient for advanced spintronics applications  \cite{jonietz2010spin,yu2012skyrmion,iwasaki2013current}. 

In bulk materials, magnetic skyrmions are commonly observed in chiral magnets, such as metallic B20 compounds (MnSi, Fe$_{1-x}$Co$_x$Si, FeGe) and insulating Cu$_2$OSeO$_3$ \cite{muhlbauer2009skyrmion,yu2010real,yu2011near,seki2012observation}. In these non-centrosymmetric compounds, a helimagnetic state is typically formed below a certain magnetic Curie temperature ($T_\mathrm{c}$) due to the competition between DM and ferromagnetic exchange interactions. In the presence of magnetic fields, a skyrmion crystal (SkX) phase emerges in the vicinity of $T_\mathrm{c}$. However, the $T_\mathrm{c}$ values of the aforementioned chiral magnets lie below room temperature, limiting the applications of skyrmions. Interestingly, this limitation is overcame in another type of chiral magnets, i.e., the $\beta$-Mn type Co$_x$Zn$_y$Mn$_z$ ($x+y+z=20$) alloys \cite{Tokunaga2015}. The Curie temperatures of Co$_x$Zn$_y$Mn$_z$ depend strongly on compositions and $T_\mathrm{c}>300$ K can be realized when $z\leq 4$ \cite{Tokunaga2015}. Above room temperature, skyrmion lattices have been observed in a series of Co$_x$Zn$_y$Mn$_z$ \cite{Tokunaga2015,kosuke2018,Yu2018,Nakajima2019,Bocarsly2019,Karube2020}. Like the B20 compounds, the SkX phase in Co$_x$Zn$_y$Mn$_z$ is typically confined in a narrow temperature ($T$) and magnetic field ($H$) region near $T_\mathrm{c}$, as a result of the interplay between long-range magnetic interactions and thermal fluctuations \cite{Tokunaga2015,kosuke2018,Yu2018,Nakajima2019,Bocarsly2019,Karube2020}. A goal in the study of skyrmions in bulk chiral magnets is to expand the temperature and magnetic field window which supports the SkX phase. 

It has been found that long-lived metastable skyrmions in Co$_x$Zn$_y$Mn$_z$ can survive to a wide temperature range well below $T_\mathrm{c}$, achieved via the assistance of magnetic frustrations, disorders, or field trained cooling \cite{Karube2016,kosuke2018,Karube2020,ukleev2021frustration}. In addition, pressure or strain is another effective tuning parameter in manipulating crystal structure, magnetic interactions and magnetic phases. It has been theoretically suggested that uniaxial anisotropy induced by uniaxial strain can stablize the skyrmion phase of MnSi, FeGe and Fe$_{1-x}$Co$_x$Si in a broad temperature and magnetic field range \cite{Butenko2010}. Indeed as found experimentally, the skyrmion phase in MnSi can be easily controlled by hydrostatic pressure \cite{Ritz2013}, uniaxial pressure \cite{Chacon2015} and uniaxial stress \cite{nii2015uniaxial}. Applying a moderate strain \cite{Seki2017} or hydrostatic pressure \cite{levatic2016dramatic} to Cu$_2$OSeO$_3$  significantly enlarges the skyrmion pocket spanned in the $T-H$ phase diagram. Pressure control of the skyrmion phase in Co$_x$Zn$_y$Mn$_z$ alloys is, however, yet to be investigated.     

In this article, we study the effects of hydrostatic pressure on the skyrmion phase in single crystals of Co$_8$Zn$_8$Mn$_4$. At ambient pressure, the skyrmion phase is found to exist within about 5 K just below $T_\mathrm{c}$. This temperature window is almost doubled under a moderate pressure of 6 kbar. The magnetic field range of the skyrmion pocket is also enlarged under pressure. Pressure is thus a sensitive tuning knob in controlling the skyrmion phase in Co$_8$Zn$_8$Mn$_4$.

\begin{figure*}
\includegraphics[width=450pt]{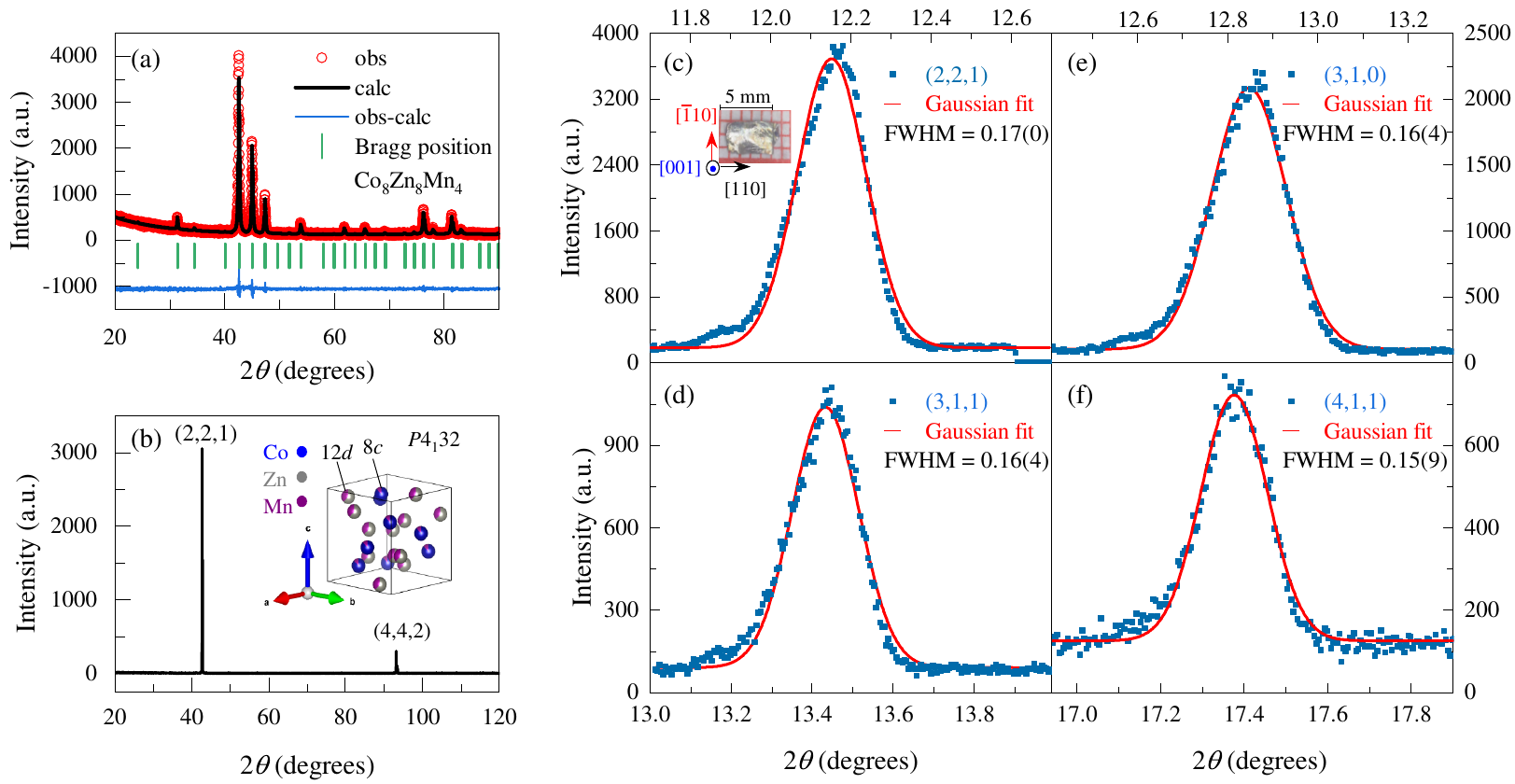}
\caption{\label{fig:1}(a) Powder X-ray diffraction pattern of the as-prepared Co$_8$Zn$_8$Mn$_4$ samples. Red circles are experimental observations. The black line is calculated diffraction data. The differences between experimental results and calculated data are presented in the blue curve. Bragg peaks are marked out by green vertical lines. (b) Single-crystal X-ray diffraction results of a Co$_8$Zn$_8$Mn$_4$ sample. The inset in (b) shows the crystal structure of $\beta$-Mn-type Co$_x$Zn$_y$Mn$_z$ alloys. (c-f) Rocking curves near four Bragg peaks measured using single-crystal neutron diffraction. The full width at half maximum (FWHM) of each peak is estimated by fitting the corresponding peak to a Gaussian profile (red line).  A photograph of a typical Co$_8$Zn$_8$Mn$_4$ single crystal is shown in the inset in (c).   } 
\end{figure*}

\section{Experimental Methods}

Single crystals of Co$_8$Zn$_8$Mn$_4$ were grown using the Bridgeman method \cite{Xie2013}. A precursor Co$_x$Zn$_y$Mn$_z$ ($x+y+z=20$) was first synthesized via a solid-state reaction. Pure Co (99.99\%), Zn (99.99\%) and Mn (99.99\%) grains were mixed in a quartz tube which was then sealed in a vacuum environment. The mixture was heated to 1000 $^{\circ}$C and kept at this temperature for 24 h. It was then slowly cooled down to 925 $^{\circ}$C at a rate of 1 $^{\circ}$C/h, followed by water quenching. Then the product was ground thoroughly in an agate mortar and the precursor complexes of Co$_8$Zn$_8$Mn$_4$ was obtained. Finally, we used these precursor complexes according to the method described in Ref. \cite{Raquet2002} to grow single crystals of Co$_8$Zn$_8$Mn$_4$. The composition of the as-prepared single crystals was determined by Energy-dispersive X-ray spectroscopy (EDX). Both powder X-ray diffraction and single-crystal X-ray diffraction (XRD) measurements were performed on a Malvern Panalytical EMPYREAN SERIES 3 diffractometer, and the latter was carried on a large piece of Co$_8$Zn$_8$Mn$_4$ single crystal (4.20 mm $\times$ 2.5 mm $\times$ 0.25 mm). Single-crystal neutron diffraction experiments were carried out on a four-circle neutron diffractometer in the Chinese Advanced Research Reactor. DC magnetization and ac magnetic susceptibility of Co$_8$Zn$_8$Mn$_4$ samples were measured in a superconducting quantum interference device magnetometer (MPMS-3, Quantum Design). DC magnetization under pressure was measured in the Magnetic Properties Measurement System (MPMS-3, Quantum Design) using a miniature piston pressure cell with Daphne 7373 as the pressure transmitting medium and Sn was loaded into a Teflon capsule along with the sample to calibrate the pressure.

\begin{figure*}
\includegraphics[width=450pt]{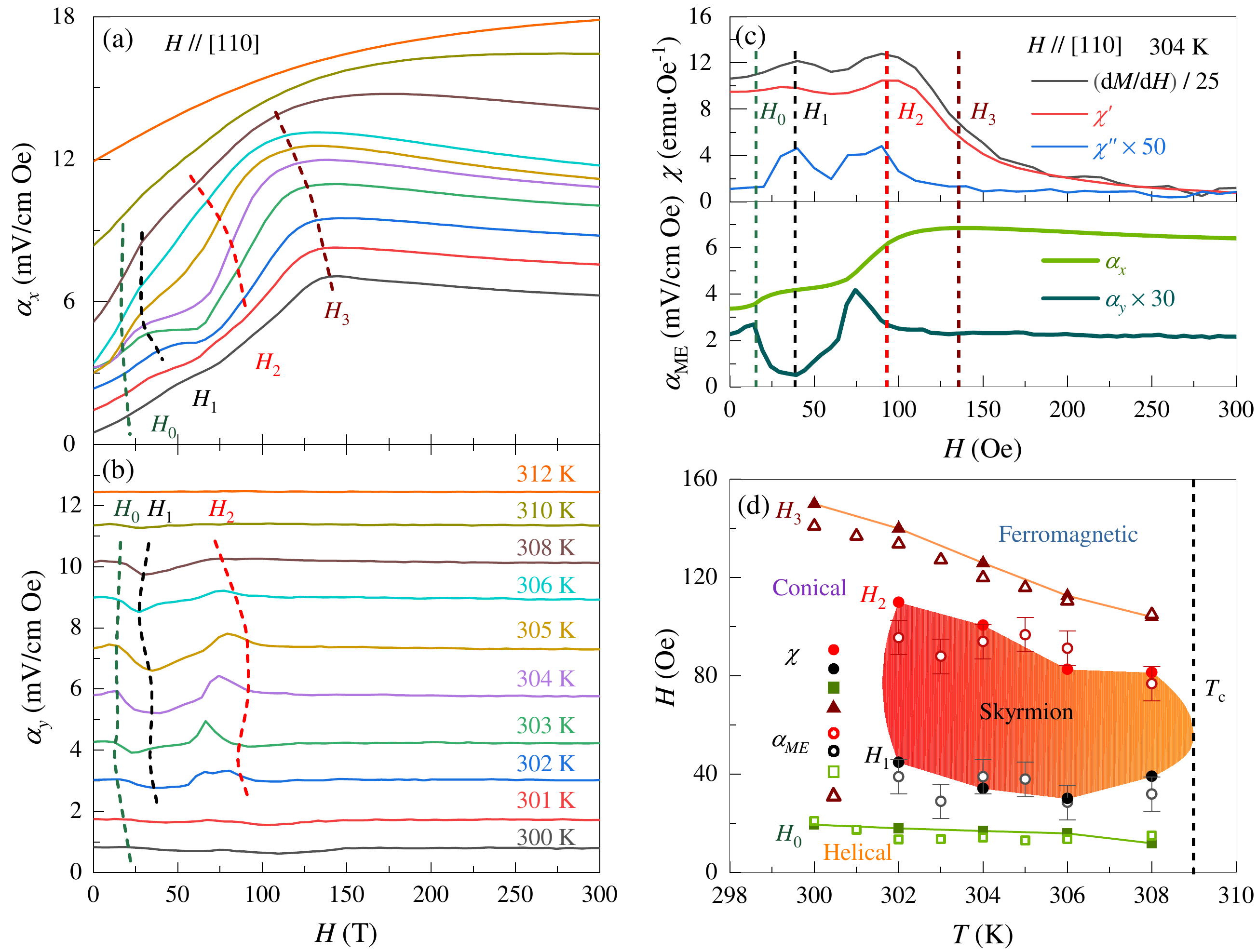}
\caption{\label{fig:3}(a) and (b) Real ($\alpha_x$) and imaginary $\alpha_y$ parts of the ac ME coefficient measured near $T_\mathrm{c}$. Curves have been shifted vertically for clarity. (c) Comparison of dc magnetization, ac susceptibility and ac ME coefficient recorded at 304 K. The $\chi''$, $\alpha_y$  and $\mathrm{d}M/\mathrm{d}H$ lines are multiplied by 50, 30 and 1/25 for better visualization. (d) $T-H$ phase diagram Co$_8$Zn$_8$Mn$_4$ obtained using ac susceptibility (solid symbols) and ac ME coefficient (open symbols) results.   }
\end{figure*}

The magnetostriction coefficient of a Co$_8$Zn$_8$Mn$_4$ single crystal was measured using a composite magnetoelectric (ME) method \cite{Chai2021, ZENG2022169631}. The Co$_8$Zn$_8$Mn$_4$ sample was mechanically bonded to a piezoelectric PMN-PT [0.7Pb(Mg$_{1/3}$Nb$_{2/3}$)O$_3$–0.3PbTiO$_3$] transducer using silver epoxy. The PMN-PT is a [001]-cut thin plate with a thickness of 0.2 mm. In a magnetic field, the magnetostrictive length deformation of Co$_8$Zn$_8$Mn$_4$ is transferred to PMN-PT, generating an electronic voltage via piezoelectric effect. The recorded ME voltage coefficient $\alpha_\mathrm{ME}$ can be formulated as :
\begin{equation}
    \alpha_\mathrm{ME}=k\frac{\mathrm{d}E}{\mathrm{d}\lambda}\frac{\mathrm{d}\lambda}{\mathrm{d}H},
\end{equation}
where $0<k<1$ is a strain transmission constant between mechanically bonded Co$_8$Zn$_8$Mn$_4$ sample and PMN-PT, $\lambda$ is the magnetostriction, $\mathrm{d}\lambda/\mathrm{d}H$ is the magnetostriction coefficient, and $\mathrm{d}E/\mathrm{d}\lambda$ is the piezoelectric coefficient. During the experiments, a small ac magnetic field was superimposed on a DC magnetic field, giving rise to an ac ME signal $\alpha_\mathrm{ME}=\alpha_x+i\alpha_y$ [see Fig. 
\ref{fig:3}]. The ac magnetic field was generated by a homemade Helmholtz coil. The ac ME data was collected using the lock-in technique.   

\begin{table}[h]
\centering
\caption{The structure parameters of Co$_8$Zn$_8$Mn$_4$ extracted from powder X-ray diffraction data measured at room temperature. The space group is $P4_132$, $a=b=c= 6.3617$ \AA, $R_\mathrm{p}$ = 5.27, $R_\mathrm{wp}$ = 9.93, $R_\mathrm{exp}=$5.91, GOF = 1.2.}
\label{tab:1}
\tabcolsep = 4 pt
\begin{tabular}{c c c c c c c }
\hline 
\hline
Atom & Wyckoff &    x    &    y    &    z    &  Biso.  &  Occ.\\
\hline 
Co(1) & 8c & 0.06477 & 0.06477 & 0.06477 & 0.521 & 0.333\\
Zn(1) & 8c & 0.06477 & 0.06477 & 0.06477 & 0.521 & 0\\
Mn(1) & 8c & 0.06477 & 0.06477 & 0.06477 & 0.521 & 0.04545\\
Co(2) & 12d & 0.125 & 0.20271 & 0.45271 & 0.703 & 0\\
Zn(2) & 12d & 0.125 & 0.20271 & 0.45271 & 0.703 & 0.3518\\
Mn(2) & 12d & 0.125 & 0.20271 & 0.45271 & 0.703 & 0.21638\\
\hline
\hline
\end{tabular}
\end{table}

\begin{figure*}
\includegraphics[width=450pt]{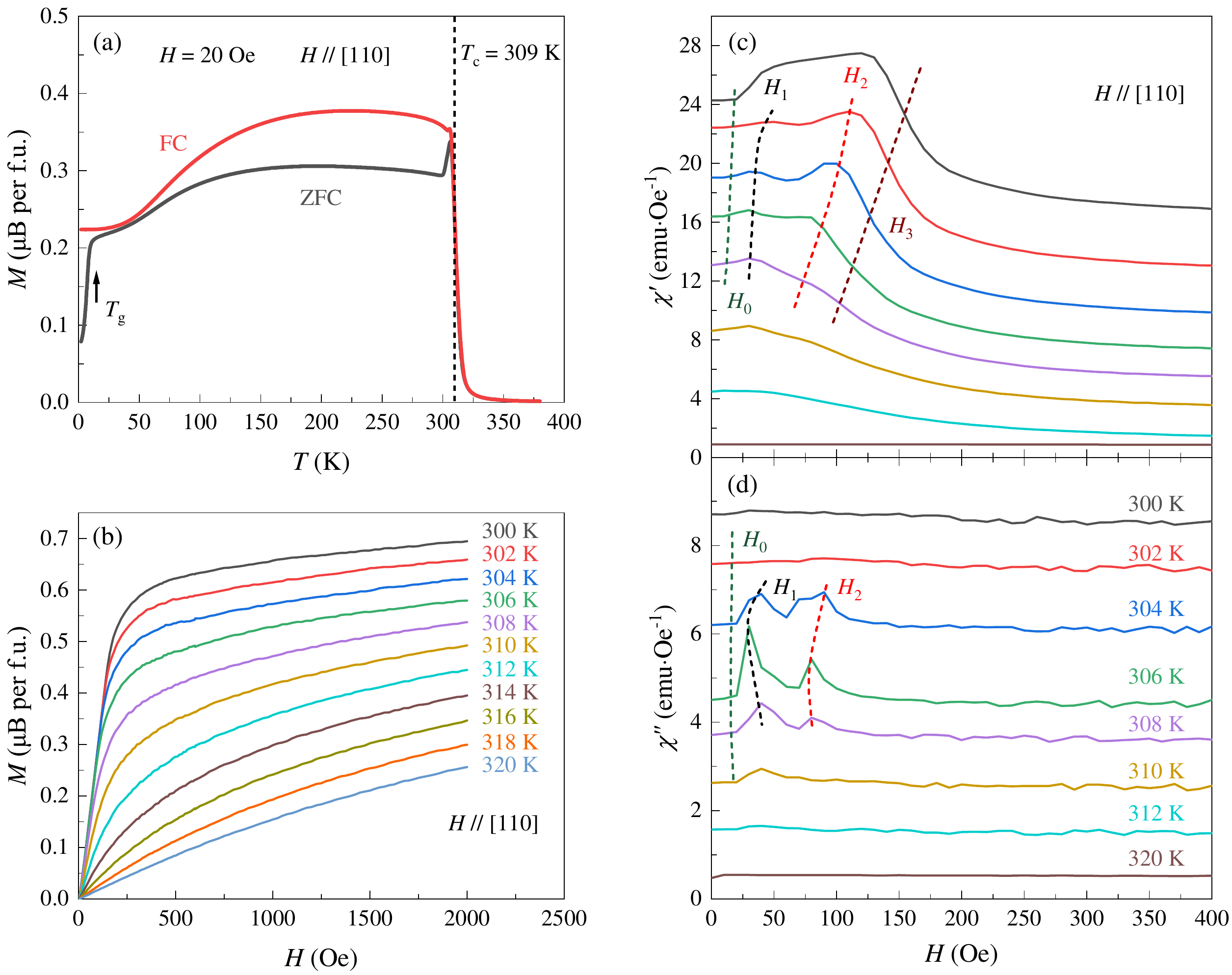}
\caption{\label{fig:2}(a) Temperature dependence of magnetization measured in a small dc magnetic field ($H=20$ Oe) applied along the [110] direction. Both zero-field-cooled (ZFC) and field-cooled (FC) data reveals a ferromagnetic-like transition at $T_\mathrm{c}=309$ K. A re-entrant spin-glass transition occurs at $T_\mathrm{g}\sim20$ K as seen in the ZFC data.  (b) Magnetization as a function of magnetic field recorded at selected temperatures near $T_\mathrm{c}$. (c) and (d) Real ($\chi'$) and imaginary $\chi''$ parts of the ac susceptibility, respectively. Curves have been shifted vertically for clarity. Green ($H_0$), black ($H_1$), red ($H_2$) and dark red (($H_3$)) lines mark the field boundaries between different magnetic phases. The skyrmion phase only exists between $H_1$ and $H_2$  }
\end{figure*}

\section{Results and Discussion}
$\beta$-Mn-type Co$_x$Zn$_y$Mn$_z$ ($x+y+z=20$) alloys crystallize in a cubic structure belonging to the chiral space group $P4_132$ [see Fig. \ref{fig:1}(b) inset] or $P4_332$ \cite{BUSCHOW19831,Xie2013}. These two types of space group host different chiralities and can be interchanged by a mirror operation. There are two different Wyckoff sites ($8c$ and $12d$) occupied by 20 atoms in each unit cell. The Co atoms are mainly coordinated at the $8c$ sties, and the Zn/Mn atoms are mostly locating at the $12d$ sites \cite{Tokunaga2015,kosuke2018,Yu2018,Nakajima2019,Bocarsly2019,Karube2020,HORI2007}. Figure \ref{fig:1}(a) shows the X-ray diffraction data obtained on Co$_8$Zn$_8$Mn$_4$ powders. The observed data can be well described by the $\beta$-Mn cubic chiral structure. The refined lattice constants give $a=b=c=6.3617$ \AA, agreeing well with previous reports \cite{Karube2016}. Detailed structure parameters are listed in Table \ref{tab:1}. It is seen that minor Mn atoms also enter the $8c$ sites, which typically happens in Co$_x$Zn$_y$Mn$_z$ compounds with $z\geq4$ \cite{Nakajima2019}. In Fig. \ref{fig:1}(b), X-ray diffraction patterns captured from a typical Co$_8$Zn$_8$Mn$_4$ single crystal are shown. Two sharp peaks can be nicely indexed by (2,2,1) and (4,4,2), respectively. The quality of the Co$_8$Zn$_8$Mn$_4$ sample is further checked by single-crystal neutron diffraction, as displayed in Figs. \ref{fig:1}(c-f). Rocking curves near four typical Bragg peaks are nicely resolved and both peaks show a very narrow full width at half maximum (FWHM) about 0.16 degrees. All these results point to the high quality of the single crystal.

\begin{figure*}
\includegraphics[width=450pt]{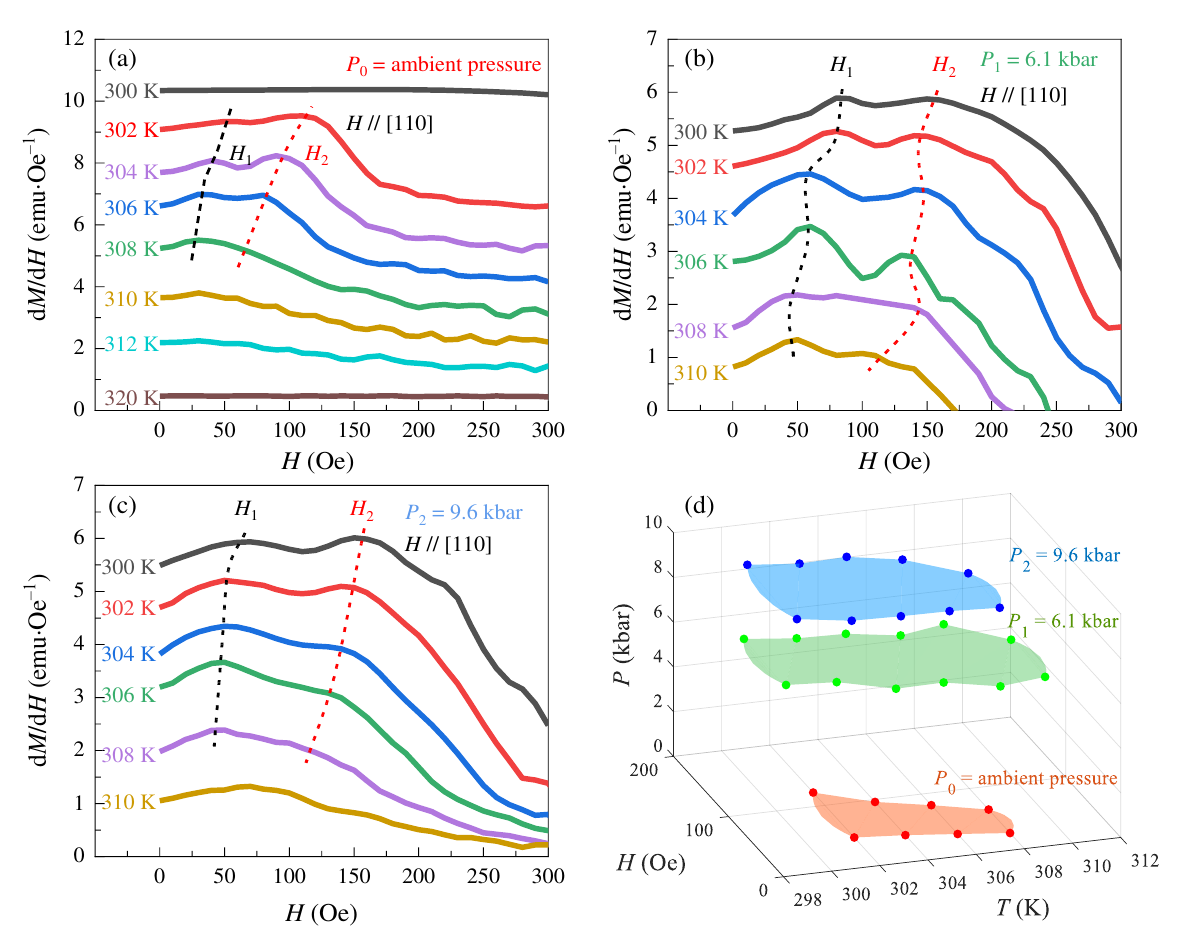}
\caption{\label{fig:4} (a-c) dc magnetization of a Co$_8$Zn$_8$Mn$_4$ single crystal measured under different pressures. Curves have been shifted vertically for clarity. (d) Comparison of the skyrmion phase captured under different pressures. The skyrmion pocket is clearly enlarged under pressure. }
\end{figure*}

Figure \ref{fig:2} shows DC magnetization data of a Co$_8$Zn$_8$Mn$_4$ single crystal.  In a weak magnetic field ($H=20$ Oe), a clear ferromagnetic-like transition is seen at $T_\mathrm{c}=309$ K in the temperature-dependent magnetization experiments [Fig. \ref{fig:2}(a)]. The transition temperature is slightly higher than that ($T_\mathrm{c}\sim300$ K) reported in other studies \cite{Tokunaga2015,Karube2016,Karube2020}. This minor difference is likely caused by slight variations in compositions, since $T_\mathrm{c}$ is sensitive to the Mn concentration \cite{Tokunaga2015,Nakajima2019,Karube2020,kosuke2018}. In zero field and below $T_\mathrm{c}$, Co$_8$Zn$_8$Mn$_4$ enters a helical magnetic state with a periodicity of $\lambda\sim125$ nm \cite{Tokunaga2015,Karube2016}. Below $T_\mathrm{c}$, the temperature dependence of magnetization deviates from a typical ferromagnet, especially at low temperatures. Below about 150 K, the magnetization deceases gradually with cooling, both in zero-field-cooled (ZFC) and field-cooled (FC) measurements. In addition, a sharp drop is seen in ZFC data below $T_\mathrm{g}\sim20$ K. The transition occurring at $T_\mathrm{g}$ is caused by a re-entrant spin-glass transition, which is commonly observed in Co$_x$Zn$_y$Mn$_z$ for $3\leq z \leq 19$ \cite{kosuke2018,Karube2020,Nakajima2019}. It has been suggested that Co moments order magnetically below $T_\mathrm{c}$, while Mn moments remain fluctuating with cooling until $T_\mathrm{g}$, below which a spin-glass state is formed \cite{Bocarsly2019,Nakajima2019}. In Fig. \ref{fig:2}(b), the magnetization near $T_\mathrm{c}$ is plotted as a function of magnetic field ($H \parallel$ [110]). Below $T_\mathrm{c}$, the magnetization rise rapidly with increasing magnetic field and become saturated above about  150 Oe. In the saturated region, magnetic moments are ferromagnetically polarized though the magnetization further increases slightly in higher fields. Slightly above $T_\mathrm{c}$, non-linear $M(H)$ curves are also seen, suggesting sizable spin fluctuations. As shown in Fig. \ref{fig:2}(a), the onset of the spin-spin correlation starts already below 330 K. And linear $M(H)$ is almost restored at 320 K [Fig. \ref{fig:2}(b)]. Note that signatures of skyrmions are masked by the rapid variations of magnetization in small magnetic fields. Transitions between different magnetic phases are seen more clearly in the derivative form $dM/dH$, as shown later in Fig. \ref{fig:4}. 

To access the skyrmion phase, we present ac susceptibility $\chi=\chi'+i\chi''$ in Figs. \ref{fig:2}(c) and \ref{fig:2}(d). Between 302 and 306 K, a clear dip is seen around 50 Oe in both the real ($\chi'$) and imaginary ($\chi''$) parts of ac susceptibility. This dip structure, commonly observed in skyrmion hosts B20 compounds and Cu$_2$OSeO$_3$ \cite{seki2012observation,Adams2012,Thessieu_1997}, is a characteristic thermodynamic signature of skyrmion lattice in Co$_8$Zn$_8$Mn$_4$ \cite{Tokunaga2015}. The magnetic fields $H_1$ and $H_2$ mark the lower and upper boundaries of the skyrmion pocket. As found by earlier reports, the skyrmion phase is sandwiched by conical states \cite{Tokunaga2015,Nakajima2019,Karube2020,kosuke2018}. The $H_1$ and $H_2$ are thus critical fields differentiating the skyrmion and conical phases. The transition between the zero-field helimagnetic and conical states happens at a lower field $H_0$, below which $\chi'$ depends weakly on $H$. In the vicinity of a higher magnetic field $H_3$, $\chi'$ drops rapidly and a polarized ferromagnetic state is achieved above $H_3$.  

In addition to the ac magnetic susceptibility, the ac ME coefficient which is proportional to magnetostriction, is another thermodynamic probe for sensing the skyrmion phase \cite{Schutte2014,Petrova2016,Chai2021,ZENG2022169631}. Distinct magnetostrictive features have been observed in the skyrmion phase of MnSi \cite{Petrova2016,Chai2021} and Co$_7$Zn$_8$Mn$_5$ \cite{ZENG2022169631}. Figures \ref{fig:3}(a) and \ref{fig:3}(b) show the ac ME coefficient results of Co$_8$Zn$_8$Mn$_4$. At temperatures above $T_\mathrm{c}$, both the real ($\alpha_x$) and imaginary ($\alpha_y$) parts of the ac ME coefficient vary smoothly with magnetic field. At 308 K just below $T_\mathrm{c}$, a kink (dip) is observed in $\alpha_x$ ($\alpha_y$) at $H_1$, pointing to the transition between conical and skyrmion phases. Upon further cooling, these features at $H_1$ become more apparent and additional signals associated with $H_2$ are also visible. The transition from conical to ferromagnetic state is visualized as a peak in $\alpha_x$ at $H_3$. Below 302 K, signatures of the skyrmion phase fade away gradually, agreeing well with the ac susceptibility data.  

In Fig. \ref{fig:3}(c), the results obtained by dc magnetization, ac susceptibility and ac ME coefficient are directly compared. It is seen that the skyrmion phase is consistently observed in all three thermodynamic probes. The critical magnetic fields associated with various magnetic transitions are also nicely identified. Based on these data, the phase diagram of Co$_8$Zn$_8$Mn$_4$ single crystals measured at ambient pressure is mapped out in Fig. \ref{fig:3}(d). Similar phase diagrams are commonly seen in Co$_x$Zn$_y$Mn$_z$ alloys \cite{Tokunaga2015,Nakajima2019,Karube2020,kosuke2018}.  In zero magnetic field, a helical magnetic state is formed below $T_\mathrm{c}=309$ K. In the magnetically ordered state just below $T_\mathrm{c}$, applying a magnetic field between $H_1$ and $H_2$ drives the system in to a skyrmion phase. Clearly, the skyrmion pocket is confined in a narrow region in the $T-H$ phase diagram.

Now we explore the effects of hydrostatic pressure on the skyrmion phase of Co$_8$Zn$_8$Mn$_4$. DC magnetization measured under different pressures are shown in Figs. \ref{fig:4}(a-c). At ambient pressure ($P_0$), the skyrmion phase is manifested as a dip in $\mathrm{d}M/\mathrm{d}H$ within $T\sim$302-306 K and $H\sim$50-100 Oe, as also found in ac susceptibility. By applying a moderate pressure of $P_1=6.1$ kbar, as presented in Fig. \ref{fig:4}(b), the dip feature in $\mathrm{d}M/\mathrm{d}H$ appears already at 310 K and persists down to 300 K. The absolute height of the dip structure is also enhanced compared to that in $P_0$. Moreover, both the lower ($H_1$) and upper ($H_2$) magnetic field boundaries of the skyrmion phase are clearly increased.  This trend persists to a higher pressure $P_2=9.6$ kbar as shown in Fig. \ref{fig:4}(c). Note that compared with $P_1$, the upper temperature limit in $P_2$ is slightly reduced back to 308 K.  Figure \ref{fig:4}(d) summarizes the phase diagram occupied by the skyrmion pocket under different pressures. Clearly, the area covered by the skyrmion phase in the $T-H$ phase diagram is significantly enlarged under pressure. Under the pressure of $P_1$, the absolute size of the skyrmion pocket is more than twice of that at ambient pressure ($P_0$). It has been shown that the skyrmion phase of Co$_8$Zn$_8$Mn$_4$ forms a triangular equilibrium lattice at ambient pressure \cite{Tokunaga2015,Karube2020,Karube2016}. The spatial organization of skyrmions formed  under pressure remains to be explored.   

At ambient pressure, the emergence of the skyrmion phase out of a conical state is necessarily stabilized by thermal fluctuations \cite{muhlbauer2009skyrmion,Buhrandt2013}. Applying pressure introduces additional parameters, which could be favorable for the stability of the skyrmion phase. Such pressure or strain tuned expansion of the skyrmion phase has been previously observed in bulk MnSi \cite{Ritz2013,Chacon2015,nii2015uniaxial}, Cu$_2$OSeO$_3$ \cite{Seki2017,levatic2016dramatic}. Using a phenomenological approach, Butenko \textit{et al.} showed that a combination of magnetic field and uniaxial anisotropy expands the skyrmion phase at the cost of the conical phase \cite{Butenko2010}.  It is now generally believed that pressure- or strain-induced magnetic anisotropy, although small, plays crucial roles in stabilizing the skyrmion phase \cite{Butenko2010,Ritz2013,Chacon2015,nii2015uniaxial,Seki2017,levatic2016dramatic}. Even in the case of hydrostatic pressure, pressure inhomogeneities and local strains may still present, leading to enhanced stability of the skyrmion phase  \cite{Ritz2013}. In addition, the DM interaction is also sensitive to pressure or strain. Enhanced DM interaction is theoretically expected along a uniaxial compressed axis \cite{koretsune2015control}. An easy-plane anisotropy is induced, which favors the skyrmion phase in a similar way to that of magnetic anisotropy. The pressure- or strain-induced modulation of the DM interaction typically leads to large deformations of the skyrmion lattice, as found in FeGe thin plates \cite{shibata2015large}. Further studies, such as small-angle neutron scattering experiments are desired to resolve the structure of the skyrmion lattice in Co$_8$Zn$_8$Mn$_4$ under pressure or strain. We also note that magnetic frustrations or disorders can induce metastable skyrmion phases in Co$_x$Zn$_y$Mn$_z$ \cite{Karube2020,kosuke2018}. While magnetic frustrations or disorders are mostly arsing from Mn atoms locating on the $12d$ sites, giving rise to metastable skyrmion phases at low temperatures. The pressure-expanded skyrmion pocket near $T_\mathrm{c}$ found in Co$_8$Zn$_8$Mn$_4$ is thus unlikely caused by  magnetic frustrations or disorders. The combined effects of thermal fluctuations, magnetic field and pressure-induced magnetic anisotropy (and/or easy-plane anisotropy) are very likely the key ingredients for stabilizing the skyrmion phase of Co$_8$Zn$_8$Mn$_4$  under pressure.

\section{Conclusions}

In summary, we have studied the evolution of the skyrmion phase in Co$_8$Zn$_8$Mn$_4$ under pressure. It is found that the size of skyrmion pocket on $T-H$ phase diagram is considerably enlarged under a moderate pressure of about 10 kbar. The enhanced stability of the skyrmion phase is likely caused by pressure-induced magnetic anisotropy and/or easy-plane anisotropy. It would be interesting to further manipulate the skyrmion phase using strain engineering on Co$_x$Zn$_y$Mn$_z$ thin films. 

\medskip

\section*{Acknowledgements}

This work has been supported by National Natural Science Foundation of China (Grants Nos. 12104254, 12227806),  Chinesisch-Deutsche Mobilit\"atsprogamm of Chinesisch-Deutsche Zentrum f\"ur Wissenschaftsf\"orderung (Grant No. M-0496), the Open Fund of the China Spallation Neutron Source Songshan Lake Science City,
the Fundamental and Applied Fundamental Research Grant of Guangdong Province (grant No. 2021B1515120015), 
the ninth batch of innovation and entrepreneurship leading talents (innovation category) in 2019, Guangdong Natural Science Funds for Distinguished Young Scholar (2021B1515020101),
Guangdong Natural Science Funds for Distinguished Young Scholar (No. 2021B1515020101), the National Key R$\&$D Program of China (No. 2023YFA1610000).

\nocite{*}

\bibliography{CoZnMn/Reference}

\end{document}